# Visually Constructing the Chemical Structure of a Single Molecule by Scanning Raman Picoscopy

Yao Zhang[†], Ben Yang[†], Atif Ghafoor[†], Yang Zhang, Yu-Fan Zhang, Rui-Pu Wang, Jin-Long Yang, Yi Luo[*], Zhen-Chao Dong[*] and J. G. Hou[*]

Hefei National Laboratory for Physical Sciences at the Microscale and Synergetic Innovation Center of Quantum Information and Quantum Physics, University of Science and Technology of China, Hefei, Anhui 230026, China.

[*]Corresponding authors: E-Mails: yiluo@ustc.edu.cn; zcdong@ustc.edu.cn; jghou@ustc.edu.cn

[†]Contributed equally to this work.

## ABSTRACT

The strong spatial confinement of a nanocavity plasmonic field has made it possible to visualize the inner structure of a single molecule and even to distinguish its vibrational modes in real space. With such ever-improved spatial resolution, it is anticipated that full vibrational imaging of a molecule could be achieved to reveal molecular structural details. Here we demonstrate full Raman images of individual vibrational modes on the Ångström level for a single Mg-porphine molecule, revealing distinct characteristics of each vibrational mode in real space. Furthermore, by exploiting the underlying interference effect and Raman fingerprint database, we propose a new methodology for structural determination, coined as scanning Raman picoscopy, to show how such ultrahigh-resolution spectromicroscopic vibrational images can be used to visually assemble the chemical structure of a single molecule through a simple Lego-like building process.

## INTRODUCTION

The determination of the chemical structure of a molecule is a premier step in chemistry. In the past century, different spectroscopic tools, such as nuclear magnetic resonance [1], electronic and vibrational spectroscopies [2–4], have been routinely employed for structure characterization. The combination of rich spectroscopic data and chemical intuitions helps to identify the basic chemical groups or specific chemical bonds in a molecule. However, the lack of spatial information has made it very difficult to firmly determine the placement and connectivity of the chemical groups from the spectroscopic data alone. Scanning tunneling microscopy (STM) [5–8] and atomic force microscopy (AFM) [9, 10] have the remarkable ability to visualize the molecular skeleton, but usually lack of sufficient chemical information required for precise chemical structure determination. Such deficiencies can in principle be overcome by a combination of scanning probe microscopy and Raman spectroscopy, as demonstrated by the tip-enhanced Raman spectroscopy (TERS) [11–25]. By taking advantage of the strong spatial confinement of the nanocavity plasmon [26–28], sub-nanometer resolution Raman images of a single molecule have been obtained, even resolving vibrational modes [22, 25, 29–31], which shows a great potential for structural determination. In this study, we present a new methodology for structural determination, named as scanning Raman picoscopy (SRP), to utilize for visually constructing the chemical structure of a single molecule. It is achieved by taking advantage of three key elements. First, the full mapping of individual vibrational modes with Ångström-level resolution allows to visually determine the placements of atoms or chemical bonds. Second, the position-dependent interference effect for local symmetric and anti-symmetric vibrations enables to identify the connectivity of the chemical groups involved. The third element is the combination of spectromicroscopic images and Raman fingerprints for different chemical groups that conclusively ensures the definite arrangement of constituent components of a single molecule. We demonstrate that the construction of a single Mg-porphine model molecule requires only a few vibrational images through a simple Lego-like building process. The protocol established in this proof-of-principle demonstration is expected to stimulate active research in the field as it develops into a mature and universal technology. To highlight the delicate structure-resolving power of this Raman-based scanning technique, the terminology scanning Raman picoscopy is adopted for such atomistic near-field tip-enhanced Raman spectromicroscopy.

**RESULTS AND DISCUSSION**

All STM imaging and Raman spectral measurements were performed on a custom-built optical-STM system operating under ultrahigh vacuum (~$5.0×10^{−11}$ Torr) and at liquid-helium cryogenic conditions (~7 K) (see Supplementary Materials S1 for more details). The SRP imaging was carried out through a synchronization function between the STM controller and CCD camera, acquiring a Raman spectrum at each pixel during scanning. As shown in the experimental setup in Fig. 1a, a single Mg-porphine model molecule adsorbed on the Ag(100) surface is excited by a confined plasmonic field generated at the apex of a Ag tip with atomic sharpness. The STM topography in Fig. 1a indicates that the molecular size is about 1 nm. Figure 1b shows typical Raman spectra for three representative positions labelled in Fig. 1a (blue, red and green dots represent for center, lobe and gap positions, respectively). Although the lateral distances among these three positions are only 3−5 Å, distinct intensity differences for different spectral peaks can already be observed. By scanning the tip over the target molecule, a series of SRP mapping images for all labelled Raman peaks are obtained (Fig. 1c), which represent the nicest vibrational images that have been experimentally observed. Each vibration shows its own characteristic image with rich details, highlighting the extraordinary power of the SRP technique. It can be estimated from the line profile of the vibrational image at 3072 cm$^{-1}$ in Fig. 1d that the spatial resolution (defined by the full width at the half maximum) can reach 1.5(1) Å (Fig. 1e). Such a high resolution enables vibrational imaging at the single-chemical-bond level. It is known that each vibrational mode is closely related to the collective motion of different atoms, which provides the information about the placement of specific atoms and their connectivity. An overall analysis of different vibrational images thus offers sufficient information for visually constructing the chemical structure of the target molecule.

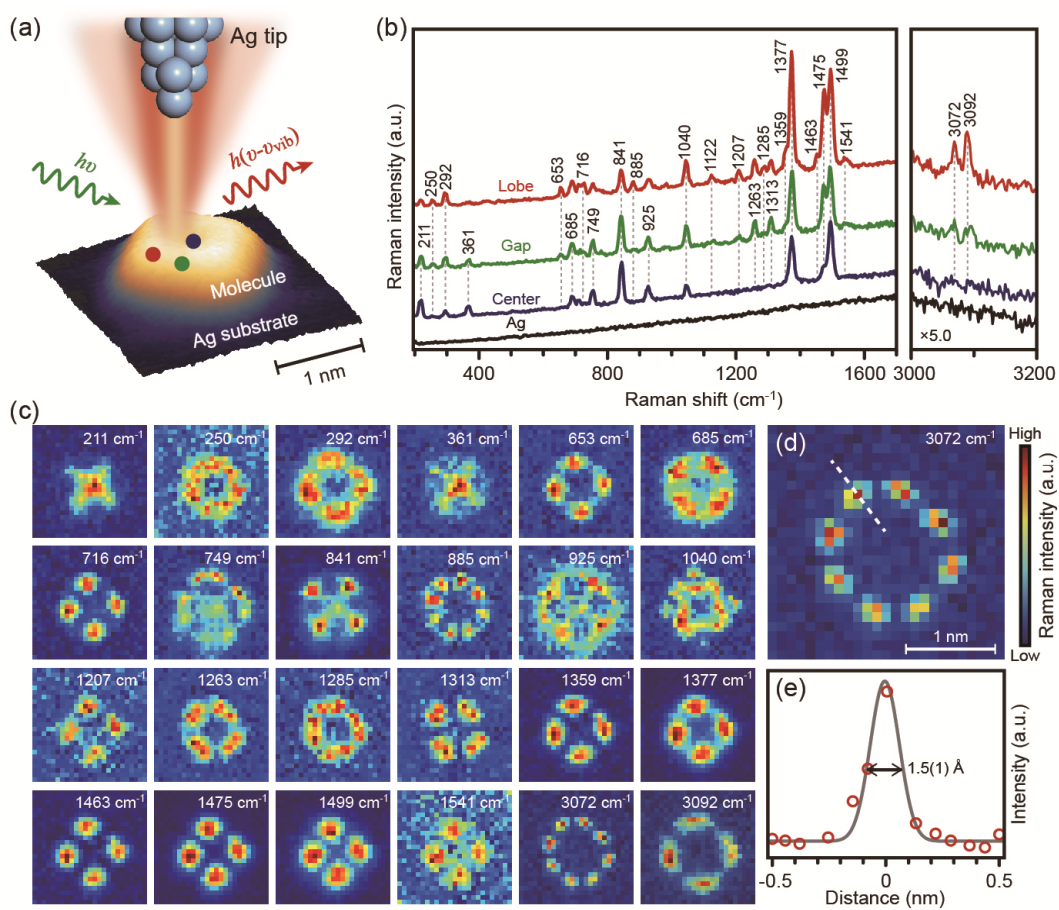

**Figure 1.** Ångström-resolved Raman images of distinct vibrational modes for a single molecule by scanning Raman picoscopy. (a) Schematic of SRP technique. The nanocavity defined by the silver tip and substrate generates a strong and highly confined plasmonic field, which is used for the excitation and emission enhancement of the Raman signals from a single molecule. The STM topograph of a single target molecule adsorbed on Ag(100) is shown at the bottom (−0.02 V, 2 pA, 2.5 nm × 2.5 nm). (b) Typical Raman spectra acquired at representative positions labelled in (a): lobe (red), gap (green) and center (blue). The spectrum on the bare Ag surface is also shown in black, confirming the clean tip condition free of contaminations. Spectral acquisition condition: −0.02 V, 8 nA, 30 s. (c) SRP spatial mapping images (−0.02 V, 8 nA, 2.5 nm × 2.5 nm, 25 × 25 pixels, 2 s per pixel) corresponding to the peaks labelled in (b), revealing different spatial distribution patterns for different Raman modes. (d) SRP

image at 3072 cm$^{-1}$ used for the estimation of spatial resolution. (e) Line profile of Raman signal intensities corresponding to the dash line in (d), exhibiting a lateral spatial resolution down to 1.5(1) Å.

The high spatial resolution of SRP images for a specific vibrational mode $Q_k$ is resulted from the confinement of the plasmonic field at the nanoscale [22, 25, 30–32]. The Raman intensity for this mode is related to the field-related vibronic transition moment between the vibronic ground ($\langle \Psi_g(Q_k,\mathbf{r})|$) and vibronic excited states ($\langle \Psi_r(Q_k,\mathbf{r})|$) (See Supplementary Materials S2 for details), i.e.,

$$\boldsymbol{\mu}_{gr}^{loc} = \int \Psi_g^*(Q_k,\mathbf{r}) \mathbf{r} g(\mathbf{r}-\mathbf{R}_0) \Psi_r(Q_k,\mathbf{r}) d\mathbf{r} \tag{1}$$

in which $g(\mathbf{r}-\mathbf{R}_0)$ is the confined field distribution function centered at the tip position $\mathbf{R}_0$. In the representation of atomic orbital basis, the wavefunctions of the ground and excited state can be described by $\Psi_g(Q_k,\mathbf{r}) = \sum_\alpha C_\alpha \varphi_\alpha(Q_k,\mathbf{r})$ and $\Psi_r(Q_k,\mathbf{r}) = \sum_\beta C_\beta \varphi_\beta(Q_k,\mathbf{r})$, respectively. Thus,

$$\begin{aligned}\boldsymbol{\mu}_{gr}^{loc} =& \sum_\alpha C_\alpha^* C_\alpha \int \mathbf{r} g(\mathbf{r}-\mathbf{R}_0) |\varphi_\alpha(Q_k,\mathbf{r})|^2 d\mathbf{r} \\ &+ \sum_\alpha \sum_{\beta \neq \alpha} C_\alpha^* C_\beta \int \mathbf{r} g(\mathbf{r}-\mathbf{R}_0) [\varphi_\alpha^*(Q_k,\mathbf{r}) \varphi_\beta(Q_k,\mathbf{r})] d\mathbf{r}\end{aligned}. \tag{2}$$

where $\varphi_{\alpha(\beta)}$ is the atomic orbital of the atom $\alpha(\beta)$. For a realistic field distribution at the Ångström level, two possible situations could occur. When the tip is on the top of an atom $\alpha$, the Raman signal from this atom becomes dominated, as represented by the first term of Eq. (2). When the tip is located at the middle of two atoms, an interference effect is expected to take place due to the cross-term $\varphi_\alpha^* \varphi_\beta$. A positive sign of this term from a symmetric vibrational motion gives a constructive signal, whereas a negative sign from an anti-symmetric vibrational motion results in a destructive signal (see Supplementary Materials S2 for details). The conceptual demonstration of the interference effect above through Eqs. (1) and (2) can be also applied to the case where two chemical bonds or multi-centers are covered by the plasmonic field. In other words, in-phase (out-of-phase) local vibrations carry the same (opposite) sign in polarization. The high spatial resolution of out-of-phase anti-symmetric vibrational modes (e.g., the 3072 cm$^{-1}$ mode in Fig. 1d) arises from the null integral under the sampling window of the local field due to the sign change.

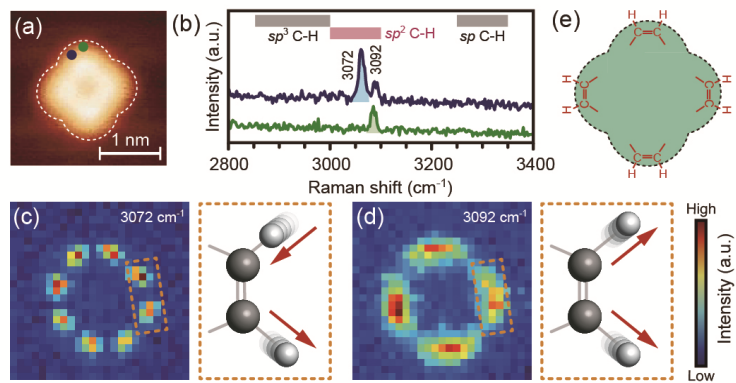

**Figure 2.** Interference effect between two neighboring $sp^2$ C−H stretching vibrations. (a) STM topograph of the target molecule (−0.02 V, 2 pA) with the dash line marking the outline of the molecule. (b) Typical single-pixel Raman spectra extracted from SRP images in high-wavenumber region acquired on slightly different lobe positions of the molecule. Blue (green) spectrum corresponds to the blue (green) dot position marked in (a). The characteristic spectral regions of different C−H stretching vibrations are marked on the top. (c, d) SRP mapping images for the two high-wavenumber Raman peaks at 3072 cm$^{-1}$ and 3092 cm$^{-1}$, showing destructive and constructive interference features respectively. The dashed boxes on the right show schematic atomic vibrations, with the red arrow pairs illustrating the anti-symmetric and symmetric vibrations. (e) Partially determined molecular structure with four H−C=C−H at the lobe positions of the molecule.

The basic information desirable for constructing a molecular structure is the types of atomic elements of the target molecule, which are known to consist of C, N, H and Mg elements for Mg-porphine. The large collection of Raman spectra for different molecules in literature and database provide tentative assumptions about the plausible functional groups involved. We start the assembling from certain well-defined Raman spectral features associated with the highly localized carbon-hydrogen (C−H) stretching vibrations in the range of 2800−3400 cm$^{-1}$ typically reported in the literature [33]. Upon slightly changing the tip position by about 0.2 nm from the blue-dot position to the more symmetric green-dot position (Fig. 2a), the Raman spectrum evolves from a two-peak feature into a single-peak feature at 3092 cm$^{-1}$ with the complete disappearance of the 3072 cm$^{-1}$ peak, as shown in Fig. 2b. Such evolution suggests the existence of two types of C−H vibrations with different characters in the small vicinity. In fact, these two vibrations can be identified as the $sp^2$ C−H stretching modes since the observed C−H stretching frequencies fall into the $sp^2$ C−H stretching region [33, 34]. The SRP images for these two modes show distinct patterns in spatial distribution, as illustrated in Fig. 2c and d. The one at 3072 cm$^{-1}$ is composed of "eight bright dots" while the other at 3092 cm$^{-1}$ consists of "four lobes", which can be well explained by the presence of the interference effect proposed above (see Supplementary Materials S2 for more details). Specifically, the well-resolved eight small dots mark the positions of eight C−H bonds, stemming from the out-of-phase destructive interference associated with an anti-symmetric vibration shown on the right of Fig. 2c (see Supplementary Video for detail). As a result, the intensities in-between the neighboring dots become the weakest. On the other hand, each lobe in the image for the 3092 cm$^{-1}$ mode is generated from the in-phase constructive interference between two neighboring C−H bonds associated with a symmetric vibration (right of Fig. 2d), which leads to the brightest spot in the lobe close to the center of two connected C−H bonds. The first Lego piece of the molecule can thus be determined to be a H−C=C−H group, and the positions of 4 pieces are illustrated in Fig. 2e.

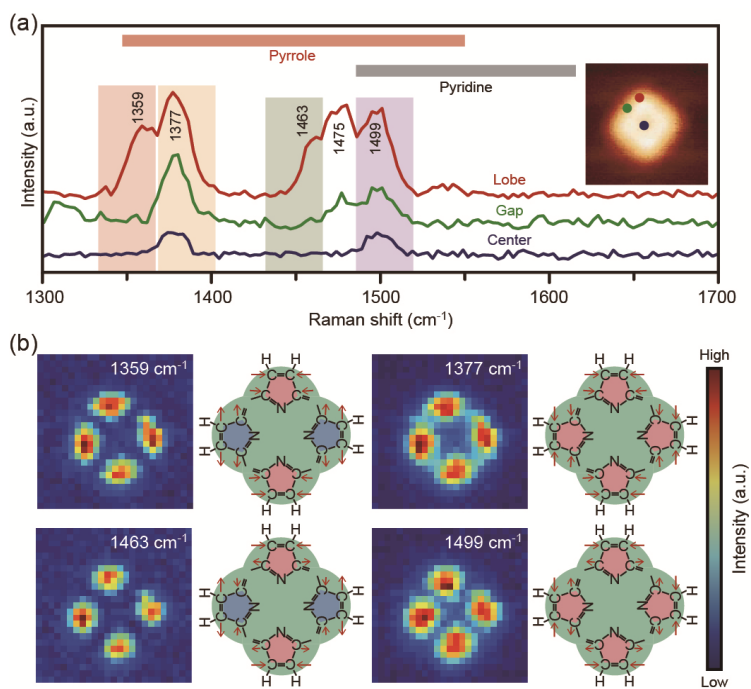

**Figure 3.** Vibrational analysis for pyrrole-ring vibrations. (a) Typical single-pixel Raman spectra in the intermediate wavenumber region acquired on the representative positions labelled in the inset STM topograph (red for lobe, green for gap and blue for center). Characteristic fingerprint regions for ring vibrations of pyrrole and pyridine are marked on the top [33]. (b) SRP mapping images for the four Raman peaks at 1359 cm$^{-1}$, 1377 cm$^{-1}$, 1463 cm$^{-1}$ and 1499 cm$^{-1}$ marked in (a). The schematic vibrations and phase relations between pyrrole rings are shown on the right of each SRP image based on the partially determined molecular structure. The pink (blue) pentagon represents the dominant "compression" ("stretching") motion of the pyrrole ring. The vibration and phase relation for the 1475 cm$^{-1}$ mode are analyzed in Supplementary Materials S3.

Next, we move on to analyze the second spectral region in 1300–1700 cm$^{-1}$ correlated with C=C stretching vibrations. The strong position dependent spectral features shown in Fig. 3a ensure a large contrast for the SRP images. Indeed, four vibrational images at 1359 cm$^{-1}$, 1377 cm$^{-1}$, 1463 cm$^{-1}$ and 1499 cm$^{-1}$ in Fig. 3b all clearly exhibit a "four-lobe" structure and each lobe has a size of about 3 Å. Considering the ultrahigh spatial resolution of 1.5 Å achieved (Fig. 1e) and the presence of $sp^2$ carbon atoms involved in C–H stretching vibrations (Fig. 2e), the absence of detailed structures within the lobe itself suggests similar Raman polarizabilities over the lobe. In other words, the electronic density over the lobe is likely to be polarized together, thus suggesting a conjugated ring structure. The SRP images in Fig. 3b for the four vibrational modes also reveal the phase relations between the vibrations of these four conjugated rings. The sharper contrast of four lobes for modes 1359 cm$^{-1}$ and 1463 cm$^{-1}$, together with negligible intensities in-between the lobes and at the molecular center, suggest a destructive interference associated with the anti-symmetric vibrational motions between neighboring rings. By contrast, the other two images at 1377 cm$^{-1}$ and 1499 cm$^{-1}$ exhibit a smaller contrast with considerable intensities between the neighboring lobes, resulting from a constructive interference associated with the symmetric vibrational motions. Moreover, the measurable intensities at the molecular center for these two symmetric modes clearly imply the presence of an atom in the center that is chemically connected to the rings (Fig. 3a). This is based on the fact that the molecular cavity size is over 4 Å, much larger than the 1.5 Å spatial resolution, consequently, the interference effect would be too small to generate the intensity in the center. With the help of Raman frequency analysis shown in Fig. 3a, one can thus conclude that the most likely structure of the ring is a five-membered pyrrole with a N atom capable of bonding to a metal. Thus, we can further build up the molecular structure by placing four pyrrole rings in the lobe positions, as shown in Fig. 3b.

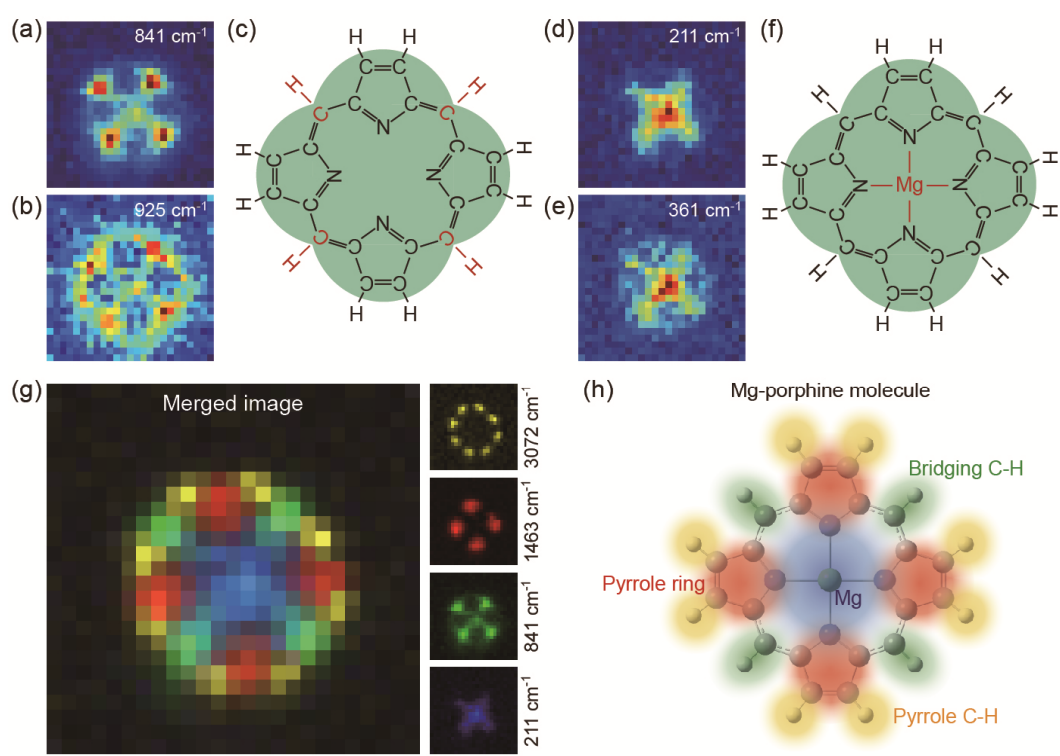

**Figure 4.** Completing full molecular structure by assembling bridging units and central metal atom. (a, b) SRP mapping images for the Raman peaks at 841 cm$^{-1}$ (a) and 925 cm$^{-1}$ (b), respectively. (c) Partially determined molecular structure including the bridging units. (d, e) SRP mapping images for the two low-wavenumber Raman peaks at 211 cm$^{-1}$ (d) and 362 cm$^{-1}$ (e), respectively. (f) Fully determined molecular structure of the Mg-porphine molecule. (g) Merged SRP image by overlaying four different image patterns shown on the right for the vibration modes at 211 cm$^{-1}$, 841 cm$^{-1}$, 1463 cm$^{-1}$ and 3072 cm$^{-1}$. (h) Artistic view of the Mg-porphine molecule showing how four colored "Legos" in (g) are assembled into a complete molecular structure, with pyrrole rings in red, pyrrole C−H bonds in yellow, bridging C−H bonds in green and central Mg atom in blue.

The next Lego piece is the one that connects the four pyrrole rings. At the vicinity of the connecting positions, the image for the vibrational mode at 841 cm$^{-1}$ shown in Fig. 4a provides the direct evidence, where four nearly isolated spots can be observed. It is known from the Raman frequency analysis, the vibrations in this frequency region is associated with C−H out-of-plane bending motions, although the C−H bending motion is not as local as the C−H stretching vibration. The possibility of a conjugated N atom acting as the bridging unit, as usually seen in porphyrazine [35], can be excluded chemically since no N−H bonds are expected for a bridging N atom, not to mention the appearance of related out-of-plane vibrations. Another support for the assignment of the bridging unit to a C−H bond is the SRP image at 925 cm$^{-1}$ for another type of C−H out-of-plane bending vibrations (Supplementary Materials S3). The brightest spot center corresponds to the bridging C−H while the elongated feature on the two sides is likely to arise from the in-phase out-of-plane bending vibrations of the two neighboring C−H on the pyrrole rings. Such a feature is not possible if the bridging unit is a N atom. The determination of the bridging "Lego" allows us to further build up a nearly complete molecular structure, which is illustrated in Fig. 4c, showing a nice porphine structure.

The last step is to determine the position of the metal atom in the center, which can be easily done by analyzing the images of low-frequency vibrations at 211 cm$^{-1}$ and 361 cm$^{-1}$, respectively, as shown in Fig. 4d and e. The large centralized spot indicates that a metal atom is chemically connected with the surrounding groups, which again confirms the assignment of pyrrole groups. The relatively large spot size suggests that the motion of the metal atom can cause wider electron density changes, beyond the porphine core area. The center atom can be assigned to Mg, since these two vibrational frequencies agree well with the Mg−N bond vibrations (Supplementary Materials S3) reported for Mg-porphine [36–38]. It should be noted that the observed SRP images in Fig. 1c all exhibit an approximate four-fold symmetry, thus ruling out the possibility of a metal-free porphine.

With the last piece in place, the chemical structure of the target Mg-porphine molecule is fully determined in real space in Fig. 4f. Moreover, the colored merged Raman image in Fig. 4g, generated by overlaying the representative individual vibrational images showing on the right side, clearly demonstrates that the spatial arrangement of individual chemical groups nicely coincides with the artistic view of the Mg-porphine molecule in Fig. 4h. The computed SRP images for the representative Raman modes that have been employed to construct the molecular structure agree very well with their experimental counterparts (see Supplementary Materials S4 for details), further confirming the experimental observation of full vibrational images and justifying the validity of the methodology proposed here.

## CONCLUSIONS

We have presented a new structural determination methodology (SRP) for visually assembling the chemical structure of a single molecule. It is achieved by combining Raman spectral fingerprints for individual chemical groups with Ångström-resolved Raman images and the interference effects involved. The rich spectral data and detailed spatial images of various vibrational modes themselves are already sufficient to provide a panoramic and global view on the molecular structure, which is more comprehensive than the verbal descriptions could give here. The Lego-like building process employed here can be easily generalized with the aid of imaging recognition and machine learning techniques, or by further combination with noncontact AFM and inelastic tunneling probe techniques. The SRP protocol established in this proof-of-principle demonstration can be widely applied for identifying the chemical structure of different materials at the level of chemical bonds.

## SUPPLEMENTARY DATA

Supplementary data are available at *NSR* online.

## ACKNOWLEDGEMENTS

We thank Prof. B. Wang for helpful discussions.

## FUNDING

This work is supported by the National Key R&D Program of China (grant numbers 2016YFA0200600 and 2017YFA0303500), the National Natural Science Foundation of China, the Chinese Academy of Sciences, and Anhui Initiative in Quantum Information Technologies.

## AUTHOR CONTRIBUTIONS

Z.C.D. and J.G.H. conceived and supervised the project. B.Y., A.G., Y.F.Z., and R.P.W. performed experiments and analyzed data. Yao Z. and Y.L. derived the theory and performed theoretical simulations. Yao Z., B.Y., A.G., Yang Z., J.L.Y., Y.L., Z.C.D. and J.G.H contributed to the data interpretation. Yao Z., B.Y., Y.L., Z.C.D. and J.G.H.

co-wrote the manuscript. All authors discussed the results and commented on the manuscript.

*Conflict of interest statement.* None declared.